\documentclass[12pt]{article}

\usepackage{graphicx}
\usepackage{epsfig}
\usepackage{amsfonts}
\usepackage{amssymb}
\newcommand{\ee}{\end{equation}}
\newcommand{\eea}{\end{eqnarray}}
\newcommand{\be}{\begin{equation}}
\newcommand{\bea}{\begin{eqnarray}}

\begin{document}

\title{
Black rings in more than five dimensions}
 \vspace{1.5truecm}
\author{ 
{\large Burkhard Kleihaus, Jutta Kunz and Eugen Radu} 
\vspace*{0.2cm}
\\
{\small Institut f\"ur Physik, Universit\"at Oldenburg,
D-26111 Oldenburg, Germany} 
   } 
\date{\today}
%\pacs{04.20.Jb, 04.40.Nr}
\maketitle

\begin{abstract} 
We construct balanced black ring solutions in $d\geq 6$ spacetime dimensions,
by solving the Einstein field equations numerically
with suitable boundary conditions.
The black ring solutions 
have a regular event horizon with $S^1\times S^{d-3}$ topology,
and they approach the Minkowski background asymptotically.
We analyze their global and horizon properties.
The Smarr formula is well satisfied.
\end{abstract}

%%%%%%%%%%%%%%%%%%%%%%%%%%%%%%%%%%%%%%%%%%%%%%%%%%%%%%%%%%%%%%%%%%%%%%%%%%%%%%
%%%%%%                     Introduction
%%%%%%%%%%%%%%%%%%%%%%%%%%%%%%%%%%%%%%%%%%%%%%%%%%%%%%%%%%%%%%%%%%%%%%%%%%%%%%

\noindent{\textbf{~~~Introduction.--~}}In 4 dimensions the stationary, 
asymptotically flat vacuum black holes (BHs) are given by the Kerr family.
A spatial section of their event horizon
has the topology of a two-sphere $S^2$.
The Kerr BHs are uniquely characterized by their
mass and angular momentum;
thus these two numbers suffice to completely specify a vacuum BH space-time.

The generalizations of the Kerr BHs to $d \geq 5$
dimensions were found by Myers and Perry (MP)
\cite{Myers:1986un}.
Their global charges are the mass and $N=[(d-1)/2]$ independent
angular momenta. Their horizon topology is that of a $(d-2)$-sphere $S^{d-2}$.
However, presenting heuristic arguments,
Myers and Perry argued 
that in higher dimensions also black rings (BRs) might exist,
and thus black objects with a different horizon topology.

In 2001 Emparan and Reall found such BR solutions
in 5 dimensions.
These are asymptotically flat and possess a
horizon topology $S^2 \times S^1$
\cite{Emparan:2001wn}.
Considering singly rotating BRs,
Emparan and Reall showed that, for fixed mass,
there are two branches of balanced black rings,
a branch of thin black rings and a branch of fat black rings.
These two branches merge 
at a minimal value of the angular momentum $j$,
where their horizon area $a_H$ exhibits a cusp.
This minimal value of $j$ of the BRs is smaller than
the maximal value of $j$ of the MP BHs.
Thus, within the range $j_{\rm min}^{BR} < j < j_{\rm max}^{MP}$,
for given global charges 3 distinct solutions exist.
Clearly, uniqueness is violated
for these 5-dimensional stationary vacuum solutions.

The discovery of the BRs spurred a lot of interest in
BH solutions in higher dimensions. With many more solutions found,
such as, in particular, composite objects of BHs and BRs,
an intriguing phase diagram of vacuum black objects
in 5 dimensions emerged
(see e.g.~\cite{Emparan:2006mm,Emparan:2008eg} 
for reviews of these aspects).
In more than 5 dimensions, however, exact solutions of BRs or 
composite black objects could not yet be obtained,
since no general analytic framework seems to exist for 
the construction of black objects
with nonspherical horizon topology in $d>5$.

A heuristic way to construct BRs is to bend 
a Schwarzschild black string and
then achieve balance by spinning it along
the $S^1$ direction \cite{Emparan:2001wn}.
This may be considered as the underlying picture for approximate techniques,
such as the method of matched asymptotic expansion 
\cite{Emparan:2007wm,Emparan:2009vd}.
Here the central assumption is that some black objects, 
in certain ultra-spinning regimes,
may be approximated by thin black strings or branes, 
curved into a given shape and boosted appropriately.

For BRs in $d>5$ dimensions this approach has led to approximate solutions,
valid for configurations with a sufficiently large radius 
of the ring \cite{Emparan:2007wm}. 
However, this approach cannot capture features
that are expected to occur at moderate values of the angular momentum,
where the radius of the $S^1$ of the ring is no longer large 
as compared to the radius of the $S^{d-3}$-sphere.
Thus, in particular, it cannot deal with the
conjectured transition of BRs to BHs with spherical horizon topology
\cite{Emparan:2007wm},
i.e., the transition to the branch of pinched black holes
emerging from a point of instability of the MP BHs
\cite{Dias:2009iu,Dias:2010maa}.

In this work we propose a nonperturbative approach to the
construction of $d>5$ vacuum BRs.
Here we obtain these solutions 
by  solving numerically the Einstein 
equations with suitable boundary conditions.
The numerical results for $d=6$
allow us to confirm the thin BR part of the phase diagram 
proposed in \cite{Emparan:2007wm}
and, in addition, to find a branch of fat BRs, which extends towards a
horizon topology changing solution.

%%%%%%%%%%%%%%%%%%%%%%%%%%%%%%%%%%%%%%%%%%%%%%%%%%%%%%%%%%%%%%%%%%%%%%%%%%%%%%%%%%
\noindent{\textbf{~~~A new coordinate system.--~}}A main difficulty 
in the construction of BR solutions is to find 
an appropriate coordinate system.
The  numerical solutions in this work are found for 
a parametrization of $D-$dimensional flat space
\begin{eqnarray}
\label{flat-space}
 ds_D^2=V_1(dr^2+r^2 d\theta^2)+V_2 d\Omega_{D-3}^2+V_3 d\psi^2,
\end{eqnarray} 
where $V_1=\frac{1}{U},~~
V_2=r^2\left(\cos^2\theta -\frac{1}{2}(1+\frac{R^2}{r^2}-U)\right),~~
V_3=r^2\left(\sin^2\theta -\frac{1}{2}(1-\frac{R^2}{r^2}-U)\right),$
with $U=\sqrt{1+\frac{R^4}{r^4}-\frac{2R^2}{r^2}\cos2\theta}$.
The coordinate range in (\ref{flat-space}) is
$0\leq r<\infty$, $0\leq \theta \leq \pi/2$,
$0\leq \psi\leq 2\pi$, 
 $d\Omega^2_{D-3}$ is the metric on the unit sphere $S^{D-3}$, 
 and $R>0$ is an arbitrary  parameter. 
 The coordinate transformation $\rho=r\sqrt{U},~\tan \Theta=(\frac{r^2+\rho^2+R^2}{r^2+\rho^2-R^2})\tan \theta ,$
leads to a more usual parametrization of the $D>3$ flat space,
$ds^2=d\rho^2+\rho^2(d\Theta^2+\cos^2\Theta d\Omega_{D-3}^2+\sin^2\Theta d\psi^2)$.

It is now manifest that for $0<r<R$, a surface of constant $r$
has ring-like topology $S^{D-2}\times S^1$, 
where the $S^1$ is parametrized by $\psi$.
The BRs will have their event horizon at a constant value of $r<R$, 
and so they will inherit this
topology\footnote{Moreover, for $d=5$, 
the $(r,\theta)$-coordinates correspond to
equipotential surfaces of a
scalar field sourced by a ring.}.

%%%%%%%%%%%%%%%%%%%%%%%%%%%%%%%%%%%%%%%%%%%%%%%%%%%%%%%%%%%%%%%%%%%%%%%%%%%%%%%%%%
\noindent{\textbf{~~~The ansatz and general relations.--~}}The metric for the $d\geq 5$ BR geometry 
preserves most of the basic structure of (\ref{flat-space}),
containing, however,  additional terms that encode the gravity effects,
 \begin{eqnarray}
\label{metric}
 ds^2=f_1(r,\theta)(dr^2+r^2 d\theta^2)+f_2(r,\theta) d\Omega_{d-4}^2+f_3(r,\theta) (d\psi-w(r,\theta) dt)^2-f_0(r,\theta) dt^2.
\end{eqnarray} 
Here the range of the radial coordinate is $r_H\leq r<\infty$, 
and $r=r_H$ corresponds to the event horizon,  
where $f_0(r_H)=0$. Thus the domain of integration has a rectangular shape,
and is well suited for numerical 
calculations.
%\footnote{Although one can devise a ring-coordinate system 
%\cite{Emparan:2006mm}
%(or more complicated versions \cite{Emparan:2007wm}) 
%for any $d > 5$, we could not use it in practice
%within a numerical scheme. 
%The main problem seems to be that spacelike infinity is approached for a
%single point in that case.}.
 
The boundary conditions satisfied at $r=r_H$ by the other metric functions are 
$2f_1+r_H\partial_{r}f_1=\partial_{r}f_2=\partial_{r}f_3=0$,   $w=\Omega_H$.
 As $r\to \infty$, the Minkowski spacetime background is recovered,
 with $f_0=f_1= 1$, $f_2=r^2\cos^2 \theta$,  $f_3=r^2\sin^2 \theta$, 
 $w=0$.
At $\theta=\pi/2$, we impose 
%\begin{eqnarray}
%\label{tpi23}
$
\partial_\theta f_0 =
\partial_\theta f_1 =
f_2 =
\partial_\theta f_3 =
\partial_\theta w=0.
$
%%%%%%%%%%%%%%%%%%%%%%%%%%%%%%%%%%%%%%%%%%%%%%%%%%%%%%%%%%%%%%%%%%%%%%%%%%%%%%
%\subsubsection{$\theta=0$, $r<R$}
%%%%%%%%%%%%%%%%%%%%%%%%%%%%%%%%%%%%%%%%%%%%%%%%%%%%%%%%%%%%%%%%%%%%%%%%%%%%%%
The boundary conditions at $\theta=0$ are 
$\partial_\theta f_0 =
\partial_\theta f_1 =
\partial_\theta f_2 =
f_3 =
\partial_\theta w =0,
$
except for the interval
$r_H< r\leq R$, where we impose instead $f_2 =\partial_\theta f_3 =0$
on the functions $f_2,f_3$.

The  metric of a spatial cross-section of the horizon is
\begin{eqnarray}
\label{eh-m}
d\sigma^2= f_{1}(r_H,\theta) r_H^2 d\theta^2 
+f_{2}(r_H, \theta)d\Omega_{d-4}^2
+f_{3}(r_H, \theta)d\psi^2.
\end{eqnarray}
From the above boundary conditions it is clear that the topology
of the horizon is $S^{d-3}\times S^1$ (although the 
$S^{d-3}$ is not a round sphere), since
$f_3$ is nonzero for any $r\leq R$, 
while $f_2$ vanishes 
at both $\theta=0$ and $\theta=\pi/2$ (which will correspond to the poles of the $S^{d-3}$-sphere).

The radii on the horizon of the ring circle, $R_1$, 
and of the $(d-3)$-sphere, $R_{d-3}$, 
are unambiguously defined only for very thin rings.
To obtain a measure for the deformation of the $S^{d-3}$ sphere,
we compare the circumference at the equator, $L_e$
($\theta=\pi/4$, where the sphere is fattest),
with the circumference of $S^{d-3}$ along the poles, $L_p$,
\begin{eqnarray}
\label{Lep}
L_e=2 \pi \sqrt{f_2(r_H,\pi/4)},~~L_p=2\int_0^{\pi/2}d\theta r_H\sqrt{f_1(r_H,\theta)},
\end{eqnarray}
and consider, in particular, their ratio $L_e/L_p$.
The radius of the $S^1$-circle
is $\theta$-dependent. 
An estimate of its deformation is given by the ratio $R_1^{(in)}/R_1^{(out)}$,
where $R_1^{(in)}$ and $R_1^{(out)}$ are the radii of the $S^1$-circle 
on the inside and outside of the ring,
respectively,
\begin{eqnarray}
\label{Ri}
R_1^{(in)}= \sqrt{f_3(r_H,0)},~~R_1^{(out)}= \sqrt{f_3(r_H,\pi/2)}.
\end{eqnarray}

A study of the $d=5$ Emparan-Reall BR written within the Ansatz (\ref{metric})  can be 
found in Ref.~\cite{Kleihaus:2010pr}, including the explicit form of the metric functions.
Note, that the $d \geq 5$ MP BH with one rotation parameter
can also be written in the form (\ref{metric}).
For BHs with a spherical horizon topology,
the metric functions satisfy the same set of boundary
conditions, except for $f_2$ and $f_3$ at $\theta=0$, 
where  $\partial_\theta f_2  = f_3 =0$ for any $r> r_H$
(see Ref.~\cite{Kleihaus:2010pr}
for a discussion of the $d=5$ case).

For both BRs and MP BHs, the event horizon area $A_H$,
Hawking temperature $T_H$ 
and event horizon velocity $\Omega_H$ of the solutions
are given by
\begin{eqnarray}
\label{eh-A}
A_H=2\pi r_H V_{d-4}
\int_0^{\pi/2}d\theta
\sqrt{f_1f_2^{d-4}f_3}\Bigg|_{r=r_H},~~
T_H= \frac{1}{2\pi}\lim_{r\to r_H} \frac{1}{(r-r_H) }\sqrt{\frac{f_{0} }{ f_{1}}},~~\Omega_H=w\big|_{r=r_H},
\end{eqnarray}
where $V_{d-4}$ is the area of the unit $S^{d-4}$ sphere. 

 The mass and angular momentum are read from the large$-r$ 
 asymptotics of the metric functions,
 $g_{tt}=-f_0=-1+\frac{C_t}{r^{d-3}}+\dots,
~g_{\psi t}=-f_3w=\sin^2\theta \frac{ C_\psi}{r^{d-3}}+\dots,$
 with ($G=1$)
\begin{eqnarray}
\label{MJ}
M=\frac{(d-2)V_{d-2}}{16 \pi }C_t,~~J=\frac{V_{d-2}}{8\pi }C_{\psi}.
\end{eqnarray} 
Also, both the MP and the BR solutions satisfy the Smarr law
\begin{eqnarray}
\label{Smarr}
\frac{d-3}{d-2}\, M=T_H\frac{A_H}{4}+\Omega_H J.
\end{eqnarray}
%and the first law $dM=T_H  {dA_H}/{4G}+\Omega_H dJ$.
%
Following \cite{Emparan:2007wm}, we define a scale by fixing the mass and introduce the dimensionless
`reduced' quantities
\begin{eqnarray}
\label{s1}
j=c_j^{\frac{1}{d-3}}  \frac{J}{M^{\frac{d-2}{d-3}}},~~
a_H=c_a^{\frac{1}{d-3}}\frac{A_H}{M^{\frac{d-2}{d-3}}},~~
w_H=c_w \Omega_H M^{\frac{1}{d-3}},~~
t_H=c_t{~}T_H M^{\frac{1}{d-3}},
\end{eqnarray}
where
%\begin{eqnarray}
%\label{coeff}
$
c_j=\frac{V_{d-3}}{2^{d+1}}\frac{(d-2)^{d-2}}{(d-3)^{(d-3)/2}}
$,
$
c_a=\frac{V_{d-3}}{2(16\pi)^{d-3}}(d-2)^{d-2}
\left(\frac{d-4}{d-3}\right)^{\frac{d-3}{2}},
$
$
c_w=\sqrt{d-3}
\left (
\frac{(d-2)}{16}V_{d-3}
\right)^{-\frac{1}{d-3}},
$
$
c_t=4\pi \sqrt{\frac{d-3}{d-4}}
\left(
\frac{d-2}{2}V_{d-3}
\right)^{-\frac{1}{d-3}}.
$
Then both the BRs and the MP BHs are conveniently parametrized by a 
single dimensionless parameter which
we choose to be $j$.

%%%%%%%%%%%%%%%%%%%%%%%%%%%%%%%%%%%%%%%%%%%%%%%%%%%%%%%%%%%%%%%%%%%%%%%%%%%%%%%%%%
\noindent{\textbf{~~~The numerical scheme.--~}}We employ a numerical 
algorithm developed in \cite{Kleihaus:2010pr,Kleihaus:2009wh} 
which uses a Newton-Raphson method to solve for $(f_i,w)$,
whilst ensuring that all the Einstein equations are satisfied.
In this approach,
the functions $f_i$ are expressed  
as products of background functions $f_i^{(0)}$ and
unknown functions $F_i$. 
For the $f_i^{(0)}$ we have chosen 
the functions corresponding to the $d=5$ static BR
(then $f_1^{(0)}$, $f_2^{(0)}$ and $f_3^{(0)}$ 
are essentially $V_1$, $V_2$ and $V_3$ in  (\ref{metric}),
though with some $r_H$-dependent corrections).
The advantage of this approach is that the coordinate singularities
are essentially subtracted,
while imposing  at the same time the event horizon topology.
The reader is referred to Ref.~\cite{Kleihaus:2010pr} 
for details of this procedure.

The equations for the $F_i$ are solved 
by using a finite difference solver and, independently, 
by using a multi-domain spectral solver,
yielding very good agreement for the results
obtained by these two different numerical schemes.
 
In our approach, 
the input parameters are $R,r_H$ and the angular velocity $\Omega_H$.
Although $R$ and $r_H$ have no invariant meaning,
they provide a rough measure of the ring's $S^1$, and of the radius of
the $S^{d-3}$ sphere, respectively,  on the horizon.
 
In 5 dimensions, BRs exist for arbitrary values of $\Omega_H$,
but these generic BRs possess conical singularities.
Only for a critical value of the event horizon velocity 
they become balanced BRs \cite{Emparan:2001wn}.
Our results indicate that for $d>5$, 
the singularities of the unbalanced configurations
are stronger. 
In this case, for given $R$, $r_H$ and arbitrary $\Omega_H$, 
the numerical algorithms do not converge well,
except for the critical value of the event horizon velocity, 
where the ring is precisely balanced.
This value is found by using a shooting procedure.
Then the balanced solution has no  singularity on and outside the horizon,
possessing a finite Kretschmann scalar.

Therefore, in principle, by varying the value of $R$
(or the position of the horizon) 
and by adjusting the value of $\Omega_H$, 
the full spectrum of $d>5$ balanced BRs
can be recovered numerically.

%%%%%%%%%%%%%%%%%%%%%%%%%%%%%%%%%%%%%%%%%%%%%%%%%%%%%%%%%%%%%%%%%%%%%%%%%%%%%%%%%%
\noindent{\textbf{~~~The results.--~}}We have studied 
in a systematic way the $d=6$ BR solutions with $1.12 r_H<R<7 r_H$.
We have used the Smarr relation (\ref{Smarr})
to test\footnote{We have also tested our numerical scheme 
by recovering the $d=5$ balanced BRs.} 
the accuracy of the numerical calculations and found it very well satisfied
with a typical relative error $<10^{-3}$. 
However, we could not obtain BRs closer to the critical point $R=r_H$
with high accuracy.

%%%%%%%%%%%%%%%%%%%%%%%%%%%%%%%%%%%%%%%%%%%%%%%%%%%%%%%%%%%
 \setlength{\unitlength}{1cm}
\begin{picture}(8,6)
\put(3,0.0){\epsfig{file=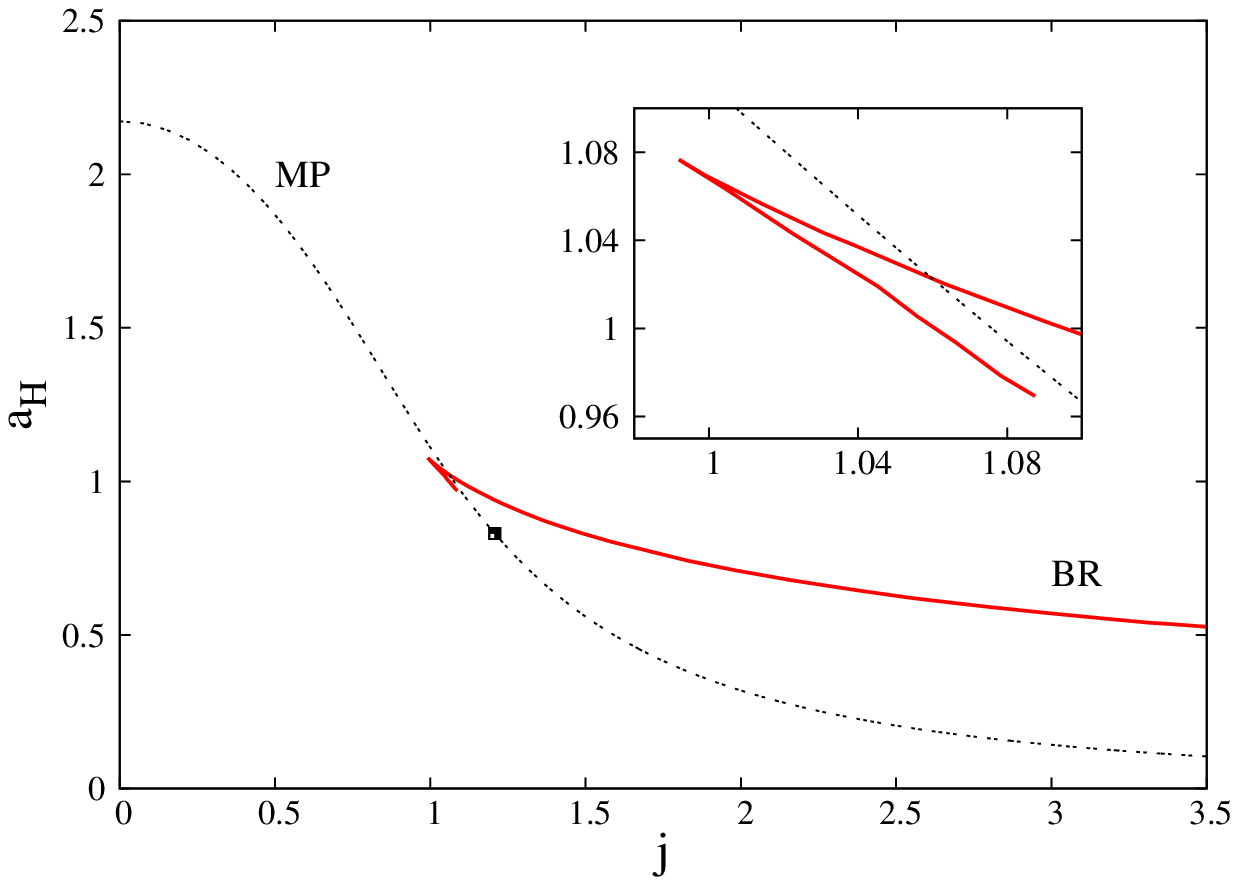,width=8cm}} 
\end{picture}
\\
\\
{\small {\bf Figure 1.}
The reduced area $a_H$ $vs.$ the reduced angular momentum $j$
for $d=6$ black rings and MP black holes with a single angular momentum.
The star on the MP curve indicates the critical solution where a branch of
`pinched' black holes should emerge
\cite{Dias:2009iu,Dias:2010maa}.} 
\vspace{0.5cm}
%%%%%%%%%%%%%%%%%%%%%%%%%%%%%%%%%%%%%%%%%%%%%%%%%%%%%%%%%%%

We have also constructed
a number of $d=7$ BR solutions, although with much larger numerical errors.
BRs should also exist for $d>7$; but so far we could not
construct them within the present scheme.
However, based on the results in \cite{Emparan:2007wm}, we expect that 
the picture found for $d=6$ is likely to hold 
for higher values of $d$ as well, thus being generic.

The profiles for the metric functions $f_i$, $w$ 
are rather similar to those of the
$d=5$ balanced BR solution exhibited in Ref.~\cite{Kleihaus:2010pr}. 
The $d>5$ BRs also possess an ergo-region, 
defined as the domain in which the metric
function $g_{tt}$ is positive.
It is bounded by the event horizon and by the surface where
$-f_0+f_3 w=0$.

The general picture we have unveiled for $d=6$ BRs 
exhibits a number of
remarkable similarities to the well-known $d=5$ case.
We again find two branches of BR solutions whose 
physical differences are most clearly seen in 
terms of the reduced quantities $a_H$ and $j$ introduced above.
The $a_H(j)$ diagram of the BRs is shown in Figure 1, with
the singly rotating MP BHs included for comparison.
The dependence of the reduced temperature $t_H$ 
and the reduced horizon angular velocity $w_H$ on the reduced
angular momentum $j$ is shown in Figure 2.

The results in Figure 1 illustrate 
that the nonuniqueness result in 5 dimensions \cite{Emparan:2001wn}
extends also to more than 5 dimensions.
We observe that the reduced area $a_H(j)$ has a cusp 
at a minimal value of $j$, $j_{\rm min}^{BR} \simeq 0.991$,
where $a_H$ assumes its maximal value, $a_H\simeq 1.076$.
Starting from this cusp the upper branch of solutions 
extends to $j\to \infty$.
In the ultra-spinning regime, these correspond to thin black rings, 
which are well approximated by boosted black strings. 
The results in \cite{Emparan:2007wm} show that a thin BR is characterized by
\begin{eqnarray}
\label{rel1}
a_H=2^{\frac{d-6}{(d-4)(d-3)}} \frac{1}{j^{\frac{1}{d-4}}},~~
w_H=\frac{1}{2j},~~
t_H=2^{-\frac{d-6}{(d-4)(d-3)}} (d-4)j^{\frac{1}{d-4}}.
\end{eqnarray} 
We have found that this analytical result provides a good approximation for $d=6$
solutions with $j \gtrsim 2$.

 %%%%%%%%%%%%%%%%%%%%%%%%%%%%%%%%%%%%%%%%%%%%%%%%%%%%%%%%%%%%%%%%%%%%%%%%%% 
\setlength{\unitlength}{1cm}
\begin{picture}(8,6)
\put(-0.5,0.0){\epsfig{file=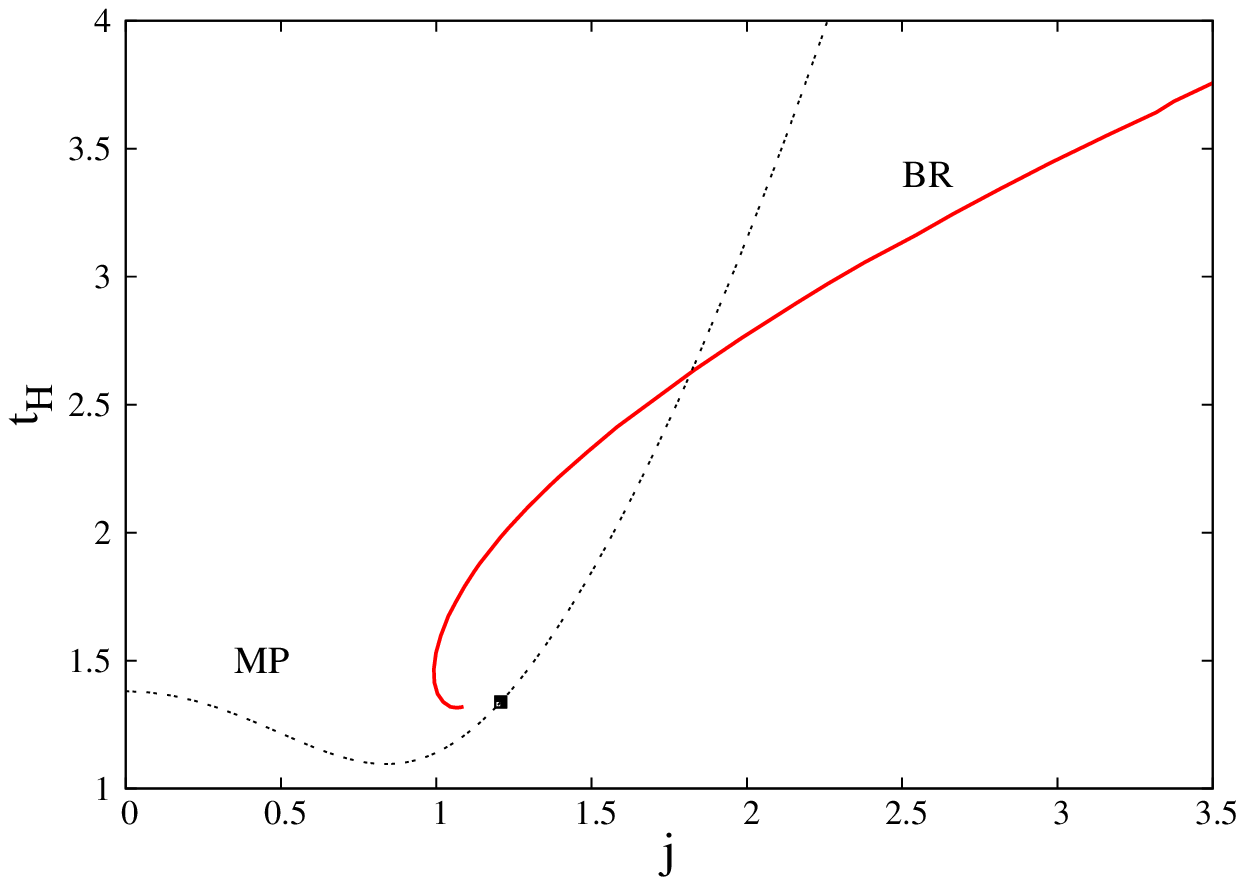,width=8cm}}
\put(8,0.0){\epsfig{file=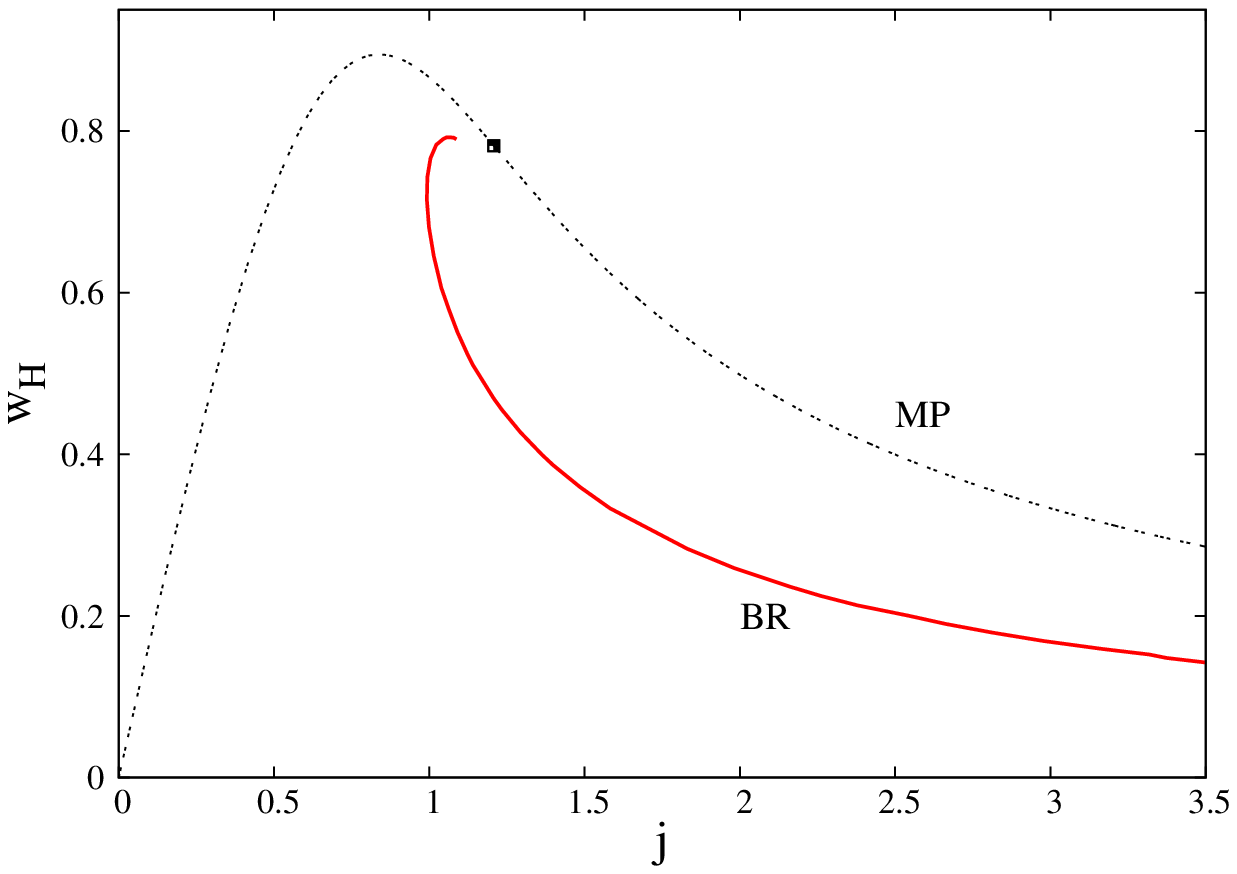,width=8cm}}
\end{picture}
\\
\\
{\small {\bf Figure 2.} The reduced temperature $t_H$
and the reduced angular velocity $w_H$ $vs.$
the reduced angular momentum $j$
for $d=6$ black rings and MP black holes.
The star on the MP curve indicates the critical solution where a branch of
`pinched' black holes should emerge
\cite{Dias:2009iu,Dias:2010maa}.
   }
\vspace{0.5cm}
%%%%%%%%%%%%%%%%%%%%%%%%%%%%%%%%%%%%%%%%%%%%%%%%%%%%%%%%%%%%%%%%%%%%%%%%%%%  

%%%%%%%%%%%%%%%%%%%%%%%%%%%%%%%%%%%%%%%%%%%%%%%%%%%%%%%%%%%
 \setlength{\unitlength}{1cm}
\begin{picture}(8,6)
\put(3,0.0){\epsfig{file=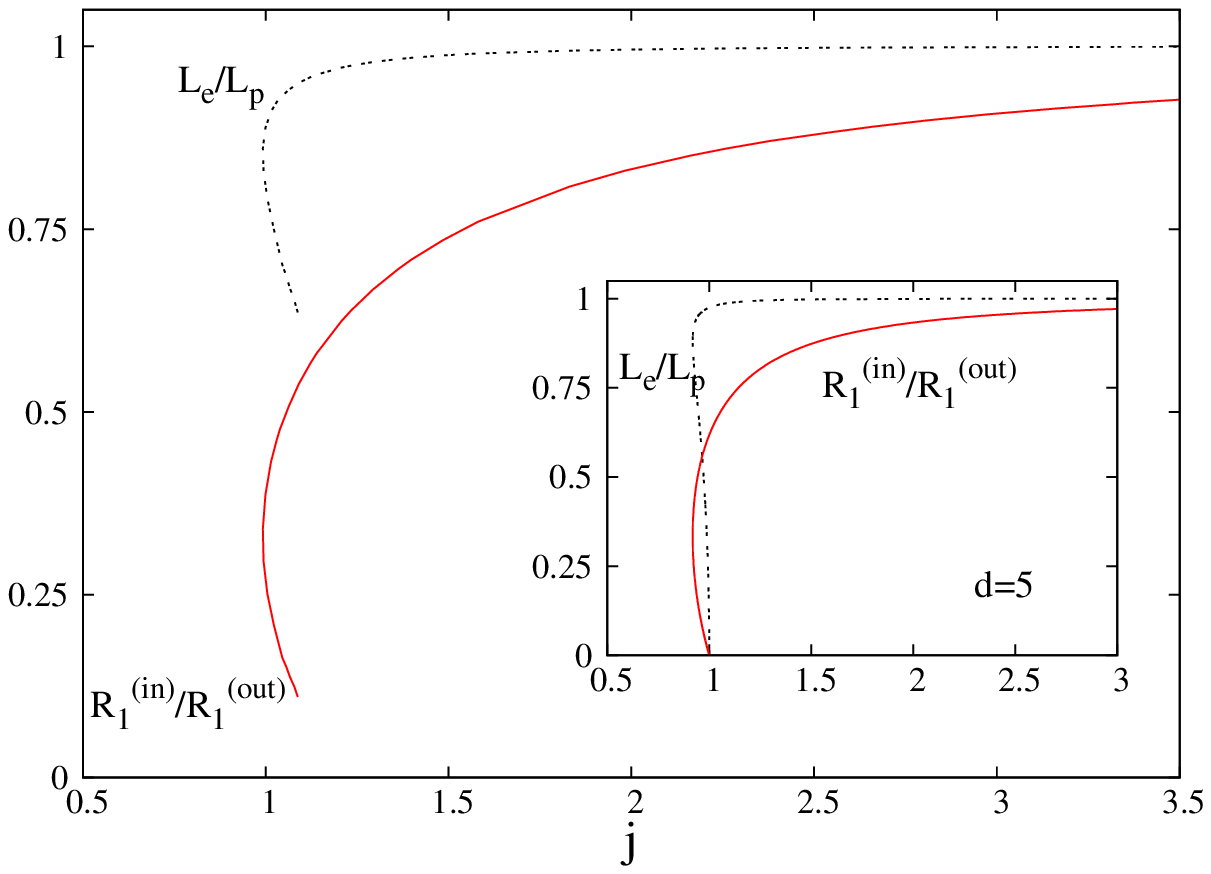,width=8cm}} 
\end{picture}
\\
\\
{\small {\bf Figure 3.}
The ratios $L_e/L_p$ and $R_1^{(in)}/R_1^{(out)}$,
which encode the deformation of the horizon,
$vs.$ the reduced angular momentum $j$ for $d=5,6$
black ring solutions. }
\vspace{0.5cm}
%%%%%%%%%%%%%%%%%%%%%%%%%%%%%%%%%%%%%%%%%%%%%%%%%%%%%%%%%%%

Starting from the cusp there is also a lower branch of BRs,
the branch of fat BRs.
Thus, as in 5 dimensions, in a certain range of the reduced
angular momentum $j_{\rm min}^{BR}< j < j_{\rm max}$
there exist three different solutions with the same global charges. 
This lower branch can only have a small extent in both $j$ and $a_H$,
since it is expected to end in a critical merger configuration
\cite{Emparan:2007wm},
where a branch of pinched black holes should be approached in a
horizon topology changing transition.
(Extrapolations of the present data
to find the location of this critical configuration
indicate that it might be in the vicinity of
$j_{max} \simeq 1.15$, $a_H \simeq 0.88$ and $t_H \sim 1.3$.)

We conclude, that the critical merger solution has a finite, nonzero $a_H$
and also $t_H$ stays finite and nonzero.
This represents a significant difference
between the $d=6$ BRs and those in $d=5$ \cite{Emparan:2001wn},
where the branch of fat BRs merges with the MP BH branch 
in a singular solution with $j=1$, $a_H=0$ and $t_H=0$.
The branch of $d=6$ `pinched' black holes itself
is expected to branch off at $j\simeq 1.27,~a_H \simeq 0.83$ 
from a critical MP BH solution, 
along the stationary zero-mode perturbation of
the GL-like instability \cite{Dias:2009iu,Dias:2010maa}.

Some of these features can be seen in Figure 3, 
where we show the deformations of the $S^1$ and $S^3$-components 
of the horizon, as defined by (\ref{Lep}) and (\ref{Ri}) 
as functions of the reduced angular momentum.
As the horizon topology changing solution is approached,
$L_e$, $L_p$ and $R_1^{(out)}$ stay finite and nonzero, 
while $R_1^{(in)}\to 0$.
In contrast, for $d=5$, 
when the hole inside the ring shrinks to zero size
the outer radius goes to infinity
as the singular solution is approached \cite{Elvang:2006dd}.
(The inset in Figure 3 demonstrates the $d=5$ case.)

%%%%%%%%%%%%%%%%%%%%%%%%%%%%%%%%%%%%%%%%%%%%%%%%%%%%%%%%%%%%%%%%%%%%%%%%%%%%%%%%%%
\noindent{\textbf{~~~Further remarks.--~}}In this work, 
by using a special coordinate system,
we have been able to formulate the problem of $d> 5$ balanced BRs 
in a numerically manageable manner, 
and to find such BR solutions in a nonperturbative way.  
Our results for $d=6$
confirm the conjecture of \cite{Emparan:2007wm}
for the phase diagram of single black objects.
Perhaps our most interesting new result is the
occurrence of a cusp in the $a_H(j)$ BR diagram
with a small branch of fat BRs.

For reasons of simplicity, 
we have restricted our calculations to vacuum 
BR solutions with a single angular momentum.
But our methods should readily extend to more general situations 
($e.g.$ to the presence of matter fields).
Moreover, it should be interesting to consider multi-black hole 
configurations and to extend 
the phase diagram proposed in \cite{Emparan:2010sx}
to the fully non-perturbative regime.
However, it currently remains a numerical challenge 
to connect the BR and MP branches 
of  $d>5$ spinning solutions by constructing the corresponding
set of `pinched' black holes.

\vspace{0.65cm}
%%%%%%%%%%%%%%%%%%%%%%%%%%%%%%%%%%%%%%%%%%%%%%%%%%%%%%%%%%%%%%%%%%%%%%%%%%%%%%%%%%
\noindent{\textbf{~~~Acknowledgements.--~}}
We would like to thank N. A. Obers for helpful discussions.
We gratefully acknowledge support by the DFG,
in particular, also within the DFG Research
Training Group 1620 ''Models of Gravity''.

%%%%%%%%%%%%%%%%%%%%%%%%%%%%%%%%%%%%%%%%%%%%%%%%%%%%%%%%%%%%%%%%%%%%%%%%%%%%%%%%%%%%%  
 \begin{small}
 
%%%%%%%%%%%%%%%%%%%%%%%%%%%%%%%%%%%%%%%%%%%%%%%%%%%%%%%%%%%%%%%%%%%%%%%%%%%%%%
 \end{small}


\begin{thebibliography}{99}
%%%%%%%%%%%%%%%%%%%%%%%%%%%%%%%%%%%%%%%%%%%%%%%%%%%%%%%%%%%%%%%%%%%%%%%%%%%%%%%%%%%%%     
%\cite{Myers:1986un}
\bibitem{Myers:1986un}
  R.~C.~Myers and M.~J.~Perry,
  %``Black Holes In Higher Dimensional Space-Times,''
  Annals Phys.\  {\bf 172}, 304 (1986).
  %%CITATION = APNYA,172,304;%%
%%%%%%%%%%%%%%%%%%%%%%%%%%%%%%%%%%%%%%%%%%%%%%%%%%%%%%%%%%%%%%%%%%%%%%%%%%%%%%%%%%%%%     
 %\cite{Emparan:2001wn}
\bibitem{Emparan:2001wn}
  R.~Emparan and H.~S.~Reall,
  %``A rotating black ring in five dimensions,''
  Phys.\ Rev.\ Lett.\  {\bf 88} (2002) 101101
  [arXiv:hep-th/0110260].
  %%CITATION = PRLTA,88,101101;%%
%%%%%%%%%%%%%%%%%%%%%%%%%%%%%%%%%%%%%%%%%%%%%%%%%%%%%%%%%%%%%%%%%%%%%%%%%%%%%%%%%%%%   
  %\cite{Emparan:2006mm}
\bibitem{Emparan:2006mm}
  R.~Emparan and H.~S.~Reall,
  %``Black rings,''
  Class.\ Quant.\ Grav.\  {\bf 23} (2006) R169
  [arXiv:hep-th/0608012].
  %%CITATION = CQGRD,23,R169;%% 
%%%%%%%%%%%%%%%%%%%%%%%%%%%%%%%%%%%%%%%%%%%%%%%%%%%%%%%%%%%%%%%%%%%%%%%%%%%%%%%%%%%%%   
 %\cite{Emparan:2008eg}
\bibitem{Emparan:2008eg}
  R.~Emparan and H.~S.~Reall,
  %``Black Holes in Higher Dimensions,''
  Living Rev.\ Rel.\  {\bf 11} (2008) 6
  [arXiv:0801.3471 [hep-th]].
  %%CITATION = 00222,11,6;%%
     
%%%%%%%%%%%%%%%%%%%%%%%%%%%%%%%%%%%%%%%%%%%%%%%%%%%%%%%%%%%%%%%%%%%%%%%%%%%%%%%
 %\cite{Emparan:2007wm}
\bibitem{Emparan:2007wm}
  R.~Emparan, T.~Harmark, V.~Niarchos, N.~A.~Obers and M.~J.~Rodriguez,
  %``The Phase Structure of Higher-Dimensional Black Rings and Black Holes,''
  JHEP {\bf 0710} (2007) 110
  [arXiv:0708.2181 [hep-th]].
  %%CITATION = JHEPA,0710,110;%%  
%%%%%%%%%%%%%%%%%%%%%%%%%%%%%%%%%%%%%%%%%%%%%%%%%%%%%%%%%%%%%%%%%%%%%%%%%%%%%%%
%\cite{Emparan:2009vd}
\bibitem{Emparan:2009vd}
  R.~Emparan, T.~Harmark, V.~Niarchos and N.~A.~Obers,
  %``New Horizons for Black Holes and Branes,''
  JHEP {\bf 1004}, 046 (2010)
  [arXiv:0912.2352 [hep-th]];
  %%CITATION = JHEPA,1004,046;%%
  %\cite{Emparan:2009cs}
\\  
%\bibitem{Emparan:2009cs}
  R.~Emparan, T.~Harmark, V.~Niarchos and N.~A.~Obers,
  %``Blackfolds,''
  Phys.\ Rev.\ Lett.\  {\bf 102} (2009) 191301
  [arXiv:0902.0427 [hep-th]].
  %%CITATION = PRLTA,102,191301; 
  

%\cite{Dias:2009iu}
\bibitem{Dias:2009iu}
  O.~J.~C.~Dias, P.~Figueras, R.~Monteiro, J.~E.~Santos and R.~Emparan,
  %``Instability and new phases of higher-dimensional rotating black holes,''
  Phys.\ Rev.\ D {\bf 80} (2009) 111701
  [arXiv:0907.2248 [hep-th]].
  %%CITATION = ARXIV:0907.2248;%%

  %\cite{Dias:2010maa}
\bibitem{Dias:2010maa}
  O.~J.~C.~Dias, P.~Figueras, R.~Monteiro and J.~E.~Santos,
  %``Ultraspinning instability of rotating black holes,''
  Phys.\ Rev.\ D {\bf 82}, 104025 (2010)
  [arXiv:1006.1904 [hep-th]].
  %%CITATION = ARXIV:1006.1904;%% 



  %\cite{Kleihaus:2010pr}
\bibitem{Kleihaus:2010pr} 
  B.~Kleihaus, J.~Kunz, E.~Radu and M.~J.~Rodriguez,
  %``New generalized nonspherical black hole solutions,''
  JHEP {\bf 1102}, 058 (2011)
  [arXiv:1010.2898 [gr-qc]].
  %%CITATION = ARXIV:1010.2898;%%
 
  
%\cite{Kleihaus:2009wh}
\bibitem{Kleihaus:2009wh}
  B.~Kleihaus, J.~Kunz, E.~Radu,
  %``d greater than or equal to five static black holes with S**2 x S**(d-4) event horizon topology,''
  Phys.\ Lett.\  {\bf B678}, 301-307 (2009).
  [arXiv:0904.2723 [hep-th]].
  
  
  %\cite{Elvang:2006dd}
\bibitem{Elvang:2006dd} 
  H.~Elvang, R.~Emparan and A.~Virmani,
  %``Dynamics and stability of black rings,''
  JHEP {\bf 0612}, 074 (2006)
  [hep-th/0608076].
  %%CITATION = HEP-TH/0608076;%% 
        
  
  %%%%%%%%%%%%%%%%%%%%%%%%%%%%%%%%%%%%%%%%%%%%%%%%%%%%%%%%%%%%%%%%%%%%%%%%%%%%%%% 
%\cite{Emparan:2010sx}
\bibitem{Emparan:2010sx}
  R.~Emparan and P.~Figueras,
  %``Multi-black rings and the phase diagram of higher-dimensional black
  %holes,''
  JHEP {\bf 1011} (2010) 022
  [arXiv:1008.3243 [hep-th]].
  %%CITATION = JHEPA,1011,022;%%
          
        
 \end{thebibliography}
\end{document}